\documentclass[useAMS,usenatbib]{mn2e}
\usepackage{graphicx,natbib,deluxetable}
\begin{document}

 \title[Planet Hunters]{Planet Hunters: The First Two Planet Candidates Identified by the Public using the Kepler Public Archive Data\thanks{This publication has been made possible by the participation of more than 40000 volunteers in the Planet Hunters project. Their contributions are individually acknowledged at \texttt{http://www.planethunters .org/authors}}} \author[Fischer et al.]{
  \parbox[t]{16cm}{Debra A. Fischer$^{2}\thanks{E-mail: debra.fischer@yale.edu}$,
Megan E. Schwamb$^{3,4,14}$,
Kevin Schawinski$^{3,4,15}$,
Chris Lintott$^{5,6}$,
John Brewer$^{2}$,
Matt Giguere$^{2}$,
Stuart Lynn$^{5}$,
Michael Parrish$^{6}$,
Thibault Sartori$^{2,7}$,
Robert Simpson$^{5}$,
Arfon Smith$^{5,6}$,
Julien Spronck$^{2}$,
Natalie Batalha$^{8}$,
Jason Rowe$^{9}$,
Jon Jenkins$^{10}$,
Steve Bryson$^{9}$,
Andrej Prsa$^{11}$,
Peter Tenenbaum$^{10}$,
Justin Crepp$^{12}$,
Tim Morton$^{12}$,
Andrew Howard$^{13}$,
Michele Beleu$^{2}$,
Zachary Kaplan$^{2}$, 
Nick vanNispen$^{2}$,
Charlie Sharzer$^{2}$,
Justin DeFouw$^{16}$, 
Agnieszka Hajduk$^{16}$,  
Joe P Neal$^{16}$, 
Adam Nemec$^{16}$, 
Nadine Schuepbach$^{16}$, 
Valerij Zimmermann$^{16}$
 \vspace{0.1in} }\\
  $^{2}$Department of Astronomy, Yale University, New Haven, CT 06511 USA\\
  $^{3}$Department of Physics, Yale University, P.O. Box 208121, New Haven, CT 06520, USA\\
  $^{4}$Yale Center for Astronomy and Astrophysics, Yale University, P.O. Box 208121, New Haven, CT 06520, USA\\
  $^{5}$Oxford Astrophysics, Denys Wilkinson Building, Keble Road, Oxford OX1 3RH\\
  $^{6}$Adler Planetarium, 1300 S. Lake Shore Drive, Chicago, IL 60605, USA\\
  $^{7}$Ecole normale superieure, 45, rue dÕUlm / 29 rue dÕUlm, F-75230 Paris cedex 05\\
  $^{8}$Department of Physics and Astronomy, San Jose State University, San Jose, CA 95192, USA\\
  $^{9}$NASA Ames Research Center, Moffett Field, CA 94035, USA\\
  $^{10}$SETI Institute/NASA Ames Research Center, Moffett Field, CA 94035\\
  $^{11}$Department of Astronomy and Astrophysics, Villanova University, 800 E. Lancaster Avenue, Villanova, PA 19085, USA\\
  $^{12}$Department of Astrophysics, California Institute of Technology, MS 249-17, Pasadena, CA 91125\\
  $^{13}$Department of Astronomy, University of California, Berkeley, CA 94720-3411\\
  $^{14}$NSF Fellow\\
  $^{15}$Einstein Fellow\\
  $^{16}$Planet Hunter\\
}

\newcommand\aj{{AJ}}%
          % Astronomical Journal
\newcommand\actaa{{Acta Astron.}}%
  % Acta Astronomica
\newcommand\araa{{ARA\&A}}%
          % Annual Review of Astron and Astrophys
\newcommand\apj{{ApJ}}%
          % Astrophysical Journal
\newcommand\apjl{{ApJ}}%
          % Astrophysical Journal, Letters
\newcommand\apjs{{ApJS}}%
          % Astrophysical Journal, Supplement
\newcommand\ao{{Appl.~Opt.}}%
          % Applied Optics
\newcommand\apss{{Ap\&SS}}%
          % Astrophysics and Space Science
\newcommand\aap{{A\&A}}%
          % Astronomy and Astrophysics
\newcommand\aapr{{A\&A~Rev.}}%
          % Astronomy and Astrophysics Reviews
\newcommand\aaps{{A\&AS}}%
          % Astronomy and Astrophysics, Supplement
\newcommand\azh{{AZh}}%
          % Astronomicheskii Zhurnal
\newcommand\baas{{BAAS}}%
          % Bulletin of the AAS
\newcommand\caa{{Chinese Astron. Astrophys.}}%
  % Chinese Astronomy and Astrophysics
\newcommand\cjaa{{Chinese J. Astron. Astrophys.}}%
  % Chinese Journal of Astronomy and Astrophysics
\newcommand\icarus{{Icarus}}%
  % Icarus
\newcommand\jcap{{J. Cosmology Astropart. Phys.}}%
  % Journal of Cosmology and Astroparticle Physics
\newcommand\jrasc{{JRASC}}%
          % Journal of the RAS of Canada
\newcommand\memras{{MmRAS}}%
          % Memoirs of the RAS
\newcommand\mnras{{MNRAS}}%
          % Monthly Notices of the RAS
\newcommand\na{{New A}}%
  % New Astronomy
\newcommand\nar{{New A Rev.}}%
  % New Astronomy Review
\newcommand\pra{{Phys.~Rev.~A}}%
          % Physical Review A: General Physics
\newcommand\prb{{Phys.~Rev.~B}}%
          % Physical Review B: Solid State
\newcommand\prc{{Phys.~Rev.~C}}%
          % Physical Review C
\newcommand\prd{{Phys.~Rev.~D}}%
          % Physical Review D
\newcommand\pre{{Phys.~Rev.~E}}%
          % Physical Review E
\newcommand\prl{{Phys.~Rev.~Lett.}}%
          % Physical Review Letters
\newcommand\pasa{{PASA}}%
  % Publications of the Astron. Soc. of Australia
\newcommand\pasp{{PASP}}%
          % Publications of the ASP
\newcommand\pasj{{PASJ}}%
          % Publications of the ASJ
\newcommand\qjras{{QJRAS}}%
          % Quarterly Journal of the RAS
\newcommand\rmxaa{{Rev. Mexicana Astron. Astrofis.}}%
  % Revista Mexicana de Astronomia y Astrofisica
\newcommand\skytel{{S\&T}}%
          % Sky and Telescope
\newcommand\solphys{{Sol.~Phys.}}%
          % Solar Physics
\newcommand\sovast{{Soviet~Ast.}}%
          % Soviet Astronomy
\newcommand\ssr{{Space~Sci.~Rev.}}%
          % Space Science Reviews
\newcommand\zap{{ZAp}}%
          % Zeitschrift fuer Astrophysik
\newcommand\nat{{Nature}}%
          % Nature
\newcommand\iaucirc{{IAU~Circ.}}%
          % IAU Cirulars
\newcommand\aplett{{Astrophys.~Lett.}}%
          % Astrophysics Letters and Communications
\newcommand\apspr{{Astrophys.~Space~Phys.~Res.}}%
          % Astrophysics Space Physics Research
\newcommand\bain{{Bull.~Astron.~Inst.~Netherlands}}%
          % Bulletin Astronomical Institute of the Netherlands
\newcommand\fcp{{Fund.~Cosmic~Phys.}}%
          % Fundamental Cosmic Physics
\newcommand\gca{{Geochim.~Cosmochim.~Acta}}%
          % Geochimica Cosmochimica Acta
\newcommand\grl{{Geophys.~Res.~Lett.}}%
          % Geophysics Research Letters
\newcommand\jcp{{J.~Chem.~Phys.}}%
          % Journal of Chemical Physics
\newcommand\jgr{{J.~Geophys.~Res.}}%
          % Journal of Geophysical Research
\newcommand\jqsrt{{J.~Quant.~Spec.~Radiat.~Transf.}}%
          % Journal of Quantitiative Spectroscopy and Radiative Trasfer
\newcommand\memsai{{Mem.~Soc.~Astron.~Italiana}}%
          % Mem. Societa Astronomica Italiana
\newcommand\nphysa{{Nucl.~Phys.~A}}%
          % Nuclear Physics A
\newcommand\physrep{{Phys.~Rep.}}%
          % Physics Reports
\newcommand\physscr{{Phys.~Scr}}%
          % Physica Scripta
\newcommand\planss{{Planet.~Space~Sci.}}%
          % Planetary Space Science
\newcommand\procspie{{Proc.~SPIE}}%
          % Proceedings of the SPIE
\newcommand\helvet{{Helvetica~Phys.~Acta}}%
          % Helvetica Phys. Acta?

\newcommand{\etal}{\mbox{\rm et al.~}}
\newcommand{\ms}{\mbox{m s$^{-1}~$}}
\newcommand{\ks}{\mbox{km s$^{-1}~$}}
\newcommand{\cse}{\mbox{cm s$^{-2}$}}
\newcommand{\kse}{\mbox{km s$^{-1}$}}
\newcommand{\mse}{\mbox{m s$^{-1}$}}
\newcommand{\msune}{M$_{\odot}$}
\newcommand{\mstar}{M$_{\star}~$}
\newcommand{\msun}{M$_{\odot}~$}
\newcommand{\lsun}{L$_{\odot}~$}
\newcommand{\rsun}{R$_{\odot}$}
\newcommand{\rstar}{R$_{\star}~$}
\newcommand{\rsune}{R$_{\odot}$}
\newcommand{\mjup}{M$_{\rm JUP}~$}
\newcommand{\mearth}{M$_{\rm EARTH}~$}
\newcommand{\msat}{M$_{\rm SAT}~$}
\newcommand{\mjupe}{M$_{\rm JUP}$}
\newcommand{\rjup}{R$_{\rm JUP}$}
\newcommand{\rearth}{R$_{\oplus}$}
\newcommand{\msini}{$M \sin i~$}
\newcommand{\vsini}{$v \sin i~$}
\newcommand{\mbsini}{$M_b \sin i~$}
\newcommand{\mcsini}{$M_c \sin i~$}
\newcommand{\chisq}{$\sqrt{\chi_{\nu}^2}~$}
\newcommand{\arel}{$a_{\rm rel}~$}
\newcommand{\teff}{$T_{\rm eff}~$}
\newcommand{\teffe}{$T_{\rm eff}$}
\newcommand{\fe}{{\rm [Fe/H]}}
\newcommand{\logg}{${\rm \log g}~$}
\newcommand{\rhk}{$\log R^\prime_{HK}~$}
\newcommand{\shk}{$S_{HK}~$}
\newcommand{\prot}{$P_{ROT}~$}
\newcommand{\caii}{{Ca}{II} H \& K}
\newcommand{\gr}{\rm g - r}
\newcommand{\snr}{\rm SNR}

%\date{Draft of 6th June 2007} edited by DT

%\pagerange{\pageref{firstpage}--\pageref{lastpage}} \pubyear{2011}

\maketitle

\label{firstpage}

\begin{abstract}
Planet Hunters is a new citizen science project, designed to engage the public in 
an exoplanet search using NASA Kepler public release data. In the first month after launch, 
users identified two new planet candidates which survived our checks for false-positives. 
The follow-up effort included analysis of Keck HIRES spectra of the host stars, analysis of pixel centroid offsets 
in the Kepler data and adaptive optics imaging at Keck using NIRC2.  Spectral 
synthesis modeling coupled with stellar evolutionary models yields a stellar density distribution, 
which is used to model the transit orbit. The orbital periods of the planet candidates 
are 9.8844 $\pm 0.0087$ days (KIC~10905746) and 49.7696 $\pm 0.00039$ (KIC~6185331) days 
and the modeled planet radii are 2.65 and 8.05 \rearth. The involvement 
of citizen scientists as part of Planet Hunters is therefore shown to be a valuable and reliable tool in exoplanet 
detection. 
\end{abstract}


\section{Introduction}
The past decade has witnessed an explosion in the number of known planets beyond our solar system. 
From the ground, planet searches using techniques that include Doppler observations, transit photometry, 
microlensing, and direct imaging have identified more than 500 exoplanets \citep{schneider11, wri11}. These 
observations have provided a wealth of information, including constraints on dynamical interactions in 
multiplanet systems, non-coplanar orbits of hot Jupiters, and atmospheric properties of transiting gas giant 
planets. The combination of Doppler and photometric measurements of transiting planets is 
particularly informative because it yields planet densities and enables theoretical modeling of the interior 
structure and composition of exoplanets.

The Kepler Mission is monitoring more than 150,000 stars with unprecedented 
29-minute observing cadence (Jenkins et al. 2010) and a relative photometric precision approaching 
20 ppm in 6.5 hours for Kp=12 mag stars to search for transiting planets. After just one year of operation, \citet{bor10a} 
announced the detection of 706 transiting planet candidates based on the first quarter 
(Q1) data.  On 2011 February 1, one month before the two-year anniversary of launch, the total 
number of planet candidates increased to more than 1200 \citep{bor11}. The Q1 data were released 
into the public archive in 2010 June, followed by a release of second quarter (Q2) data 
in 2011 Februrary.  The public archive is hosted by the Multi-mission Archive at STScI (MAST\footnote{http://archive.stsci.edu/}) 
and the NASA/IPAC/NExSci Star and Exoplanet Database (NStED\footnote{http://nsted.ipac.caltech.edu}).

Although there are more than 1200 Kepler candidates, only $1 - 2$\% of these are 
confirmed planets with measured masses from Doppler observations \citep{bat11, bor10b}.
These are challenging confirmations. The Kepler 
stars are faint compared to stars in ground based radial velocity surveys and most of 
the Kepler candidates have radii consistent with Neptune like planets, so most of the 
stellar reflex velocities are comparable to the formal measurement errors. Transit timing 
variations \citep{hol10, liss11} offer a novel way to derive planet masses, but require multi-planet systems with 
measureable non-Keplerian orbital perturbations. 

The Kepler team has developed sophisticated algorithms for detecting transits 
by fitting and removing periodic or quasi-periodic stellar variability (with low and high 
frequencies). In addition to modeling out background variability, the Kepler pipeline stitches 
together data from different observing quarters by determining the median flux from adjacent 
observing windows and using polynomial fits across the boundary. The Kepler team developed the 
Transit Planet Search (TPS) algorithm, a wavelet-based adaptive filter to identify a periodic 
pulse train with temporal widths ranging from 1 to 16 hours \citep{jenkins02, jenkins10}.
Photometric uncertainties are assessed to identify light curves with phase-folded detection 
statistics exceeding 7.1-sigma. This threshold was selected so that given the number
of required independent statistical tests per star, four years of data for the entire set of Kepler targets 
could be robustly searched for orbital periods up to two years.  

While the human brain is exceptionally good at detecting patterns, 
it is impractical for a single individual to review each of the $\sim150,000$ light curves 
in every quarterly release of the Kepler database. However, crowd-sourcing this task has 
appeal because human classifiers have a remarkable ability to recognize archetypes and to 
assemble groups of similar objects, while disregarding obvious glitches that can trip up 
computer algorithms.  This skill has recently been 
put to use in a wide range of scientific fields, from galaxy morphology to protein folding. To 
engage these uniquely human talents, and to give the public the opportunity to participate in an 
exciting exoplanet search, we developed Planet Hunters\footnote{www.planethunters.org} 
to present Kepler light curves to the public.

Planet Hunters is a new addition to the successful Zooniverse network of Citizen 
Science Alliance projects \citep{lin08, lin11}, and the first Zooniverse project to present 
time series data (rather than images) to the public. The site was launched on 
2010 December 16, and after six months, more than 40,000 users have made more than 
3 million light curve classifications. Here we describe the layout of the site and 
two new planet candidates identified by the public using the PlanetHunter interface. 

\section{Identifying Transits}
The Planet Hunters website makes use of the Zooniverse\footnote{www.zooniverse.org} 
toolset, which now supports a wide variety of citizen science projects. Its primary function 
is to serve up ÔassetsÕ - in this case $\sim$33 day flux-corrected light curves derived from the Kepler data - to 
an interface, and to collect user-generated interactions with these data.

Previous Zooniverse projects have included a separate tutorial to assist volunteers. 
While the Planet Hunters website includes such a tutorial, initial guidance is given within 
the interface, accessed via a single click from the site home page. Volunteers see a light 
curve with example transits, and can then begin to classify data. Users who have not 
registered with the Zooniverse, or who are not logged in, can begin classifying but receive 
frequent reminders to log in. The site supports prioritization of the light curves; for logged-in 
users viewing the Q1 data discussed in this paper, simulated or already identified transits 
were shown 5\% of the time.  A curve associated with a dwarf star was then shown 
66\% of the remaining time, and one associated with a giant star 33\% of the time. 
Once a category (i.e., simulated light curve, dwarf or giant star) has been selected, a 
light curve is chosen randomly from the top ten scoring assets in that category. (The 
score is the number of transits marked on each curve). Once curves have been classified 
by ten volunteers, they are removed from 
the list. The results are made available to the science team immediately via a private website. 

\begin{figure*}
\begin{center}
\includegraphics[angle=0, width=1.0\textwidth]{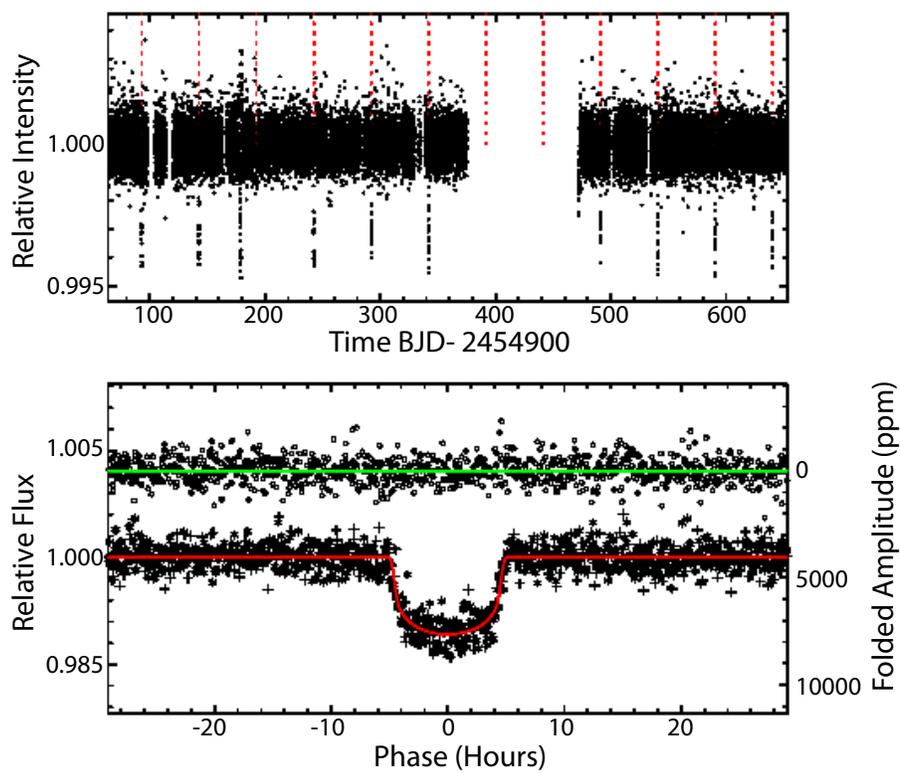}
\caption{These slides from the Planet Hunters interface show the light curve for KOI 889.01 (top). Participants use a mouse-drag to identify prospective transit features (bottom).}

\label{fig_interface_figure}

\end{center}
\end{figure*}

The actual classification proceeds via a decision tree. In the first step, users are 
asked whether the light curve is variable or quiet (icons and help buttons provide 
visual prompts).  The user is then 
asked whether any transit features are present and has the option to zoom in and out of particular 
areas of the light curve. If transit features are found, the user can mark them with 
boxes as demonstrated in Figure \ref{fig_interface_figure}. In some cases, the transit features 
seen are synthetic transits of known period and radius, which are used to assess the completeness 
of the user classifications.  

After all transits are marked, the user has the option to discuss this particular star 
on the Planet Hunters Talk site and connect with other citizen scientists. The user 
can also download the light curve data to analyze it independently or save the star to 
their ``favorites''.  The Discussion Board (``Talk''\footnote{http://talk.planethunters.org/}) 
is a critical component of the Planet Hunters project.  Here, 
the science team interacts with the public and experienced users establish collections 
of similar light curves (e.g., ``Variables in a Hurry,''  ``Definite Transits,'' ``Weird Stars'') and 
provide advice for new users. The integration of discussion into the workflow has been 
successful in encouraging greater participation than in previous Zooniverse projects; more 
than 60\% of registered Planet Hunters participants visit "Talk," and more than 35\% make 
comments.

\subsection{Planet Hunters Detection Efficiency}
As a first check, we visually inspected all user assessments made in the first month 
after the site was launched for the first 306 Kepler planet candidates announced 
by \citet{bor10a}. This essentially provided a ``head count'' or a rough estimate of how many transit 
events were being flagged by participants and it provided feedback that was considered by the web 
development team for upgrades to the site (e.g., streamlining the assessment questions and transit
marking routines). Note that this is simply a tally of the fraction of transits that were marked; we are not 
calculating the percentage of planets detected. For example, if a sample of ten stars had one hundred transit 
events and 80 of them were marked by 50\% of classifiers, then the percentage of detected transits 
would be 0.8*50 = 40\%. The 306 Kepler planet candidates \citep{bor10a}, exhibited 1371 transits 
with planet radii between 0.1 and 1 \rjup. Overall, we found that two thirds of the transits for candidates 
announced by \citet{bor10a} were correctly flagged. Only 10\% of transit boxes were spurious (i.e., did 
not obviously correspond to a transit event). 

\section{Kepler Planet Hunters Candidates} 
We also visually inspected $\sim 3500$ transit flags marked by Planet Hunters in light curves where five or more 
people indicated that a transit had been found. We first eliminated the 
known false positives, typically grazing and eclipsing binaries \citep{bat10, prsa10, row10}, and published Kepler 
candidates \citep{bor10a, bor11} from the set of light curves flagged by Planet Hunters.  On our 
internal web site, the team searched the extracted light curves, ran periodogram analyses, 
modeled light curves for prospective candidates and checked for correlated pixel 
brightness centroid shifts to try to eliminate additional false positives. After an 
extensive filtering process, we reduced the number of possible planet candidates down 
to a preliminary list of forty five.  

We ranked these candidates and sent the ``top ten'' to our Kepler co-authors; they examined the 
light curves with their data verification pipeline and immediately found that six of the ten were 
unlikely to be planet candidates. KIC~11904734 has a V-shaped transit and very large radius, 
suggesting an eclipsing binary star system. KIC~8043052 and KIC~12009347 have 
secondary occultations that are also consistent with eclipsing binary systems. 
KIC~4913000 and KIC~9097892 showed changing transit depths 
from quarter to quarter. This can occur when a nearby star contributes an amount of flux that 
is quarter dependent, changing as the instrumental point spread function changes. 
A more complete pixel centroid analysis showed that the transit signals for KIC~4913000, 
KIC~8242434, and KIC~9097892 were offset 
from the star by $4 - 6$ arcseconds. KIC~11820830 initially appeared to be a strong planet 
candidate, however stellar modeling indicated that the most likely interpretation for this star was 
that it was an eclipsing binary (EB) system with a large early type star as the primary and a M or K dwarf
secondary. The six false positive candidates are listed in Table \ref{tab_failed}.

However, three candidates survived the Kepler data verification pipelines. One of these is a 
possible multi-planet candidate and we are now obtaining Doppler follow-up. The remaining 
two candidates are presented here. Each of these candidates had in fact been flagged 
in Q1 by the Kepler TPS as Threshold Crossing Events. However, for various reasons, 
these objects were not promoted to the status of a ``Kepler Object of Interest,'' or KOI.   

\subsection{KIC~10905746\label{sec_10905746}} 
KIC~10905746 has a Kepler magnitude of 13.496 and \gr\ color of 0.949. The Kepler Input Catalog \citep{kmt09} 
does not list \teffe, \logg, \fe\ or stellar radius for this star. The star was dropped 
from the Kepler target list after Q1 because variability characteristics 
(amplitude and frequency) indicated that the star could be a giant 
and was therefore less desirable for the exoplanet transit survey; planet transit signals are 
much shallower and more difficult to detect around stars with large radii. The photometry for this star 
shows low frequency variability, with a period of $\sim16$ days and an amplitude of more than 2\%, 
which could be caused by spots rotating on the surface of the star. 

The Planet Hunters participants were able to look past the large scale structure in the light curve 
and they identified possible transit events with a depth of about 0.2\% that repeated on $\sim10$-day intervals 
in the Q1 data. The shape and depth of the light curve seemed consistent with a planet and we did 
not detect photocenter offsets in the pixel arrays in our initial screening, which would have indicated a blended background 
eclipsing binary system. 

\begin{figure*}
\begin{center}
\includegraphics[angle=0, width=1.0\textwidth]{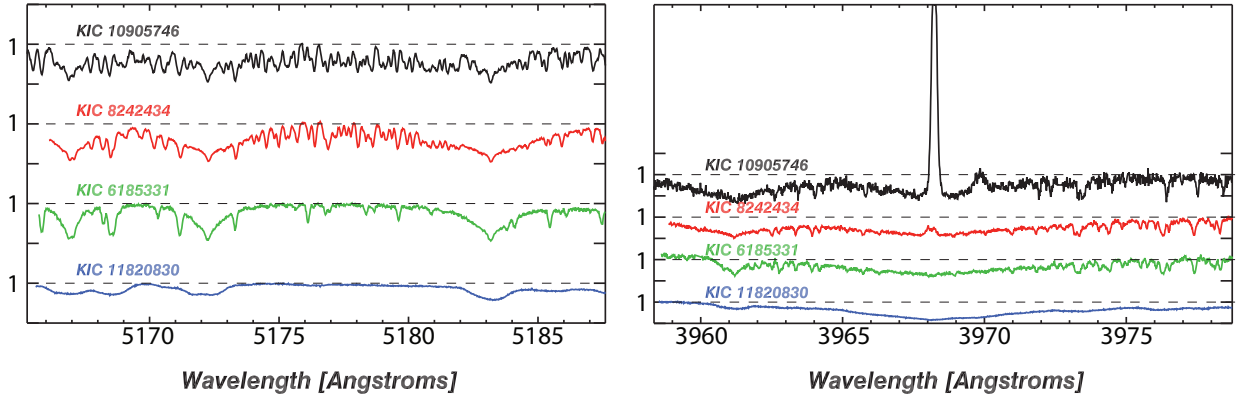}
\caption{(left) The wings of the Mg B triplet lines are sensitive to pressure broadening, making 
these lines useful diagnostics of the surface gravity or luminosity class of stars. The spectra above 
were obtained at Keck and the stars are ordered from high to low surface gravity based on our 
spectral synthesis models. (right) Emission in the cores of the \caii\ line is an activity indicator for 
main sequence stars. The spectra above show the Ca II K line for each of the planet candidate 
hosts presented here. The strong emission for KIC 10905746 is typical for a late-type main 
sequence star. }

\label{fig_mgb_caii}

\end{center}
\end{figure*}

To better understand the host star, we obtained a spectrum of this star at Keck with resolution 
of $R \sim 55000$, using  HIRES \citep{vogt94} on 2011 April 12.  A faint companion was observed at 
a separation of about 5" on the guide camera and the image rotator was used to ensure that the light 
from the companion did not enter the slit.  With the excellent seeing and the greater than one magnitude 
difference between KIC~10905746 and the companion star, the scattered light contamination would have been less 
than one part in a thousand. The spectrum had a signal-to-noise 
ratio of about 140 and we used the Spectroscopy Made Easy (SME) code \citep{vp96, vf05} to model the 
stellar parameters: \teff = 4237 $\pm 114$K, \logg= 4.73$\pm 0.1$, \vsini $= 1.1 \pm 1$ \ks, and 
\fe\ $= -0.23 \pm 0.1$.  The surface gravity that we measure with our LTE spectroscopic analysis 
is consistent with a main sequence star, rather than an evolved giant. Figure \ref{fig_mgb_caii} (left, top row) shows a 
wavelength segment that includes the Mg I B triplet lines from the Keck spectrum. The wings of 
these lines are sensitive gravity indicators. However, in this case, the star is cool with significant line blanketing, 
which suppresses the continuum and makes it difficult to model the line wings.
We tested the hypothesis that this star was a giant by running a grid of synthetic models and 
fixing the gravity between \logg of 2.0 - 3.5. The chi-squared fit for our models improved with 
decreasing surface gravity over this range, but all fits were significantly worse than our model 
with \logg = 4.73. 

The \caii\ lines provide additional support of main sequence status for this star. 
Late type main sequence stars often have significant emission in the spectral line cores 
as a result of dynamo-driven magnetic activity in the star, like the strong emission in 
the \caii\ line cores, shown in Figure \ref{fig_mgb_caii} for KIC~10905746.  However, it is far less common 
for evolved stars to show emission unless the stars are rapidly rotating or members of 
close spectroscopic binary systems \citep{if10, giz02, gun98, gn85}, and we see no evidence for either of these 
attributes in KIC~10905746.  The combination of emission in the cores of the \caii\ and pressure-broadened 
wings in the Mg I B lines, together with the spectroscopic \teffe, suggests 
that the star has a spectral type of roughly M0V.
The stellar parameters are summarized in Table \ref{tab_stellar_pars} .

Our Kepler co-authors found that the Kepler TPS algorithm had flagged the 
light curve for KIC~10905746 in Q1 with a Multiple Event Statistic 
(MES) of $9 \sigma$, greater than the $7.1 \sigma$ threshold. However, the fit failed to 
converge during the next stage of data verification. As a result, the star was dropped, the full pipeline analysis was 
never carried out until it was flagged by the Planet Hunters. 

\begin{figure*}
\begin{center}
\includegraphics[angle=0, width=1.0\textwidth]{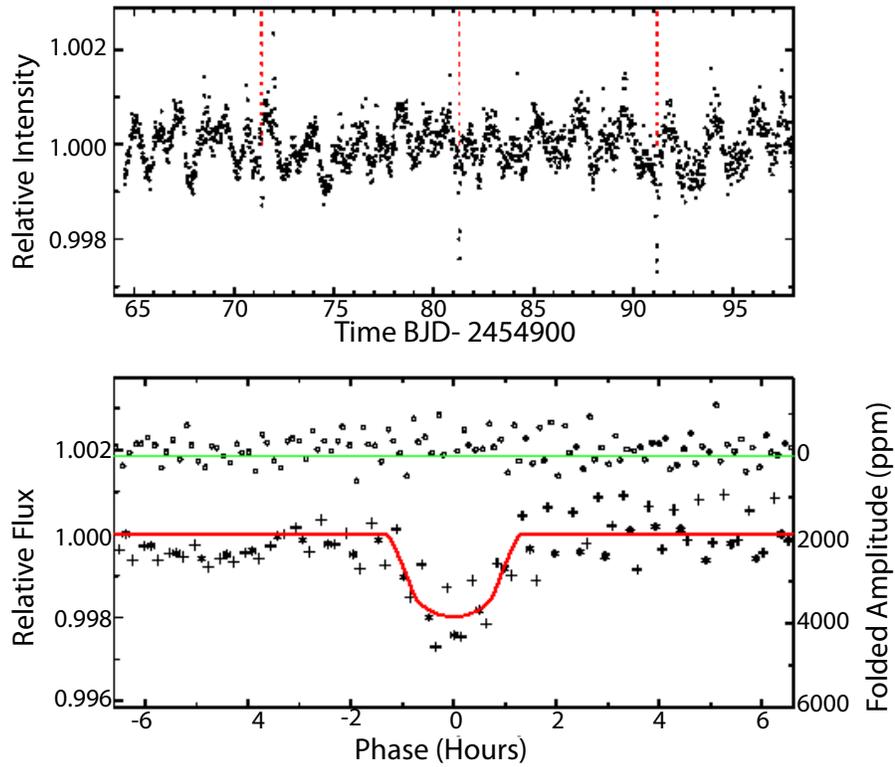}
\caption{The top panel shows the time series data for KIC~10905746 between 2009 May 2 and 
2009 June 15 after removing a large amplitude periodic signal. Planet Hunters flagged the three 
transit events indicated with a vertical dashed red line in the Q1 data. 
In the bottom panel, the light curve is phase-folded at the prospective orbital period 
P = 9.8846 days after removing the baseline variability. The fitted transit model is 
overplotted with a red curve. Just above the transit light curve, the anti-transit photometry
is plotted and fit with a green curve showing zero depth for the occultation.}

\label{fig_10905746_lc}

\end{center}
\end{figure*}

The Kepler time series photometry for Q1 is shown in the top panel of Figure \ref{fig_10905746_lc} 
(after removing the large amplitude, low frequency variability). The bottom panel of 
Figure \ref{fig_10905746_lc} shows the data folded at the prospective orbital period and the 
red curve is the best fit theoretical curve with a period 
of $9.8844 \pm 0.0087$d, an orbital inclination of 88.42 degrees and an inferred 
planet radius of 2.65 $\pm 0.67$\rearth.  Just above the transit curve, we show the 
photometry from the opposite phase, where a putative secondary occultation might 
be observed.  The search for secondary occultations allowed for eccentric orbits that were 
consistent with the data and stellar parameters. The anti-transit data are folded at a phase of 0.5
since no eccentricity or secondary occultations were detected when modeling the light curves.

A Monte Carlo analysis  \citep{jen08} 
iterates between a family of evolutionary models in the Yale-Yonsei 
isochrones \citep{d04, yi04}, and the spectroscopic parameters and orbital 
parameters (orbital period, transit depth and duration) to provide 
self-consistent estimates for uncertainties and stellar parameters, including 
Z (total heavy element abundance), age, density, luminosity, 
mass and radius. For KIC~10905746, age is not listed in Table \ref{tab_stellar_pars} since there 
was almost no constraint from the evolutionary tracks. 
Since the transit depth is a function of the ratio of the planet to star radius, 
an accurate assessment of the stellar radius is critical for deriving the planet 
radius. The characteristics of this planet candidate 
are summarized in Table \ref{tab_planet_pars}. 

Because we do not have an independent measurement of the mass of the transiting 
object, KIC10905746 is a planet candidates rather than a confirmed planet. 
Photometrically diluted background eclipsing binaries (BGEB) can have 
transit depths similar to planets. The depth of an eclipsing binary system will normally be 
10\% or more (depending on the ratio of the stellar radii and the impact parameter), but if the 
eclipsing binary light curve is blended with a brighter foreground star, the 
composite light curve will have a shallower depth during the eclipse and can 
masquerade as a transiting planet candidate.  However, 
other signatures of the BGEB can sometimes be found 
in the light curve: unequal primary eclipse and secondary occultation 
or V-shaped light curves \citep{bat10}.  Three tests were 
carried out to search for a BGEB. First, the light curve was examined for 
deviations from a planet model (e.g., variations in the depths of 
alternating transits or evidence for secondary occultations). 
In Figure \ref{fig_10905746_lc}, the even and odd transits are indicated with plus symbols and asterisks 
respectively and show that the alternating transit events do not have significant variations 
in depth and are well-fit with a transiting planet model, which is overplotted as a solid
line. The photometric data plotted just above the transit curve are phase-folded 
at the predicted time of secondary occultation for a BGEB and fit with a theoretical 
(green) line that solves for an occultation with zero depth.  We note that 
the search for occultations does not assume zero eccentricity, however, 
zero eccentricity is used to generate the anti-transit phased plot. 
For many BGEB's some dimming would be 
observed. The lack of a detected occultation is a necessary, but still not 
exclusive condition for a planet origin of the transit event. 

To place stronger limits on the presence of a blended BGEB, 
adaptive optics (AO) observations were obtained on 2011 June 23 UT 
using NIRC2 at Keck. The conditions were excellent with $\sim$0\farcs5 seeing 
and very little cirrus. The spatial resolution of the K-band AO images is about 45~mas. 
Figure \ref{fig_im_montage} (top panel) compares a K-band image of KIC 10905746 from 
2MASS\footnote{http://irsa.ipac.caltech.edu/applications/FinderChart} (left) 
with our diffraction-limited K-band AO images (right) with square root scaling for the brightness. 
The 2MASS image is unresolved, but 
reveals a faint source $\sim$4\farcs2 east of KIC~10905746, identified as 
KIC~10905748. The high resolution K-band AO images cleanly resolves these 
two sources. Our ability to rule out other close companions depends on the brightness
contrast of the stars in K-band and their angular separation. 
The $3 \sigma$ magnitude differences for excluding other sources are listed in 
Table \ref{tab_ao} for separations ranging from 0\farcs25 out to 4\farcs0.  We also 
obtained J-band images to better characterize the neighboring source. The 
magnitude difference between KIC~10905746 and KIC~10905748 is 
$\Delta K = 1.42$ mags and $\Delta J = 1.38$ mags. These 
images did not reveal any additional prospective contaminating sources.

The file headers of the Kepler data contain information about the pixel centroid 
at the time of every photometric measurement.  If the transit is occurring on the source, 
then the brightness of the star will decrease, but the image centroid position will be unchanged.  
However, if we are really observing a blended system with a background eclipsing binary
that is offset from the source, then the image centroid will shift during the eclipse.   
Centroids for the pixel images in the Kepler data were examined for this  
astrometric motion. The pixel centroid analysis yielded a high SNR detection for 
KIC~10905746 and no sign of astrometric motion was detected beyond the 
error circle of beyond 0.08 pixels. While these results do not rule out a 
background binary close to KIC~10905746, they do eliminate the nearby star, KIC~10905748, 
which is $46 \sigma$ away in the model fit, as the source of the transit. 

\subsection{KIC 6185331}
The Planet Hunters identified a single transit event for KIC~6185331 in the Q1 data and 
one additional transit was found in the Q2 Kepler light curve.  The Kepler team notes that the TPS 
code also identified this as a prospective candidate with a MES of about $10 \sigma$ in Q1 
and $20 \sigma$ in Q2. However, the data verification pipeline did not trigger to process 
these curves. 

According to the Kepler Input Catalog, KIC~6185331 has a Kepler magnitude of 
15.64, \gr\ color of 0.556, stellar radius of 0.664 \rsun, \teff = 5578, \logg= 4.786, 
and \fe\ =-0.287.  We obtained a spectrum of this star with \snr $\sim 30$ 
using HIRES at Keck Observatory.  Our spectral synthesis modeling with SME 
yields an effective temperature of $5615 \pm 80$K, consistent with the KIC value.  
However, our analysis yields a lower gravity of \logg = 4.19 $\pm 0.15$.
Comparing the Mg I B triplet lines (Figure \ref{fig_mgb_caii}) there is indeed less pressure 
broadening than for KIC~8242434, which had a log g of 4.608.
We also derive a slightly less metal rich composition than the KIC, with 
$\fe\ = +0.11 \pm 0.1$ and we obtain a best fit model for the lines with \vsini = 0.5 \kse. No 
emission is seen in the core of the \caii\ lines (Figure \ref{fig_mgb_caii}), indicating that this sunlike star 
has low chromospheric activity. 

\begin{figure*}
\begin{center}
\includegraphics[angle=0, width=1.0\textwidth]{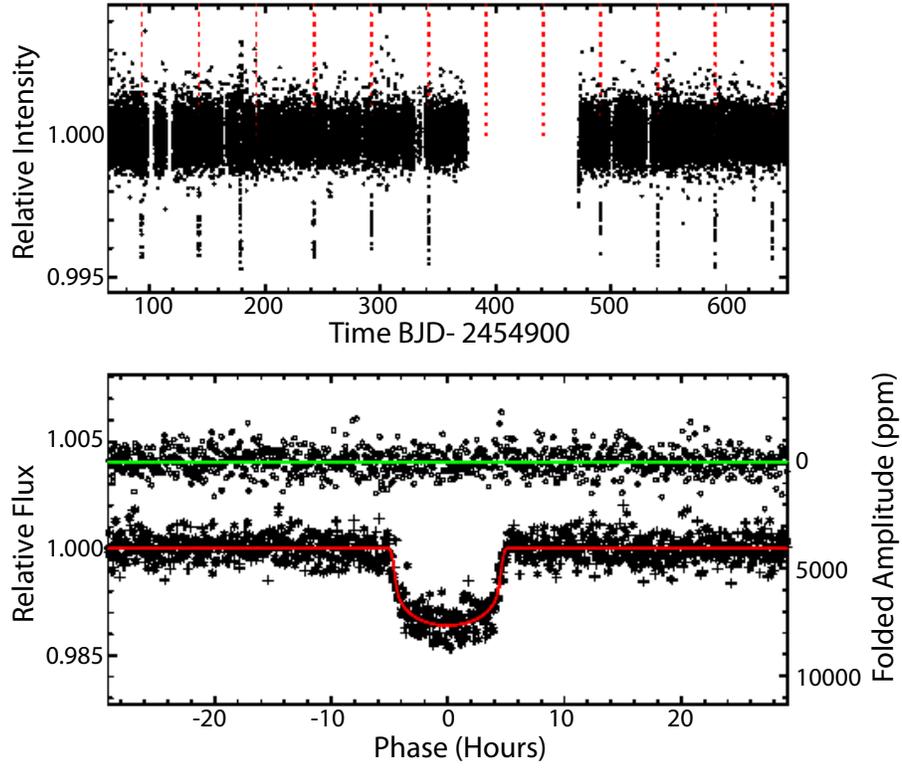}
\caption{The time series data for KIC~6185331 (top) include Q1 - Q7 data. Planet Hunters flagged 
a single transit in the Q1 data and a one additional 
transit was seen in the Q2 data.  The bottom panel shows the data folded at the prospective  
orbital period, 49.7700 days.}

\label{fig_6185331_lc}

\end{center}
\end{figure*}

Figure \ref{fig_6185331_lc} shows the time series data (top) and the phase-folded data 
(bottom), modeled with a 49.76971d period using the Q1 - Q7 data.  
We carried out the Monte Carlo analysis described in \S\ref{sec_10905746} for KIC~10905746 
with the Y2 isochrones, orbital parameters and the spectral synthesis results to obtain self-consistent 
stellar parameters (listed in Table \ref{tab_stellar_pars}). With the derived stellar radius of 
1.27 \rsun, the planet is modeled with a best fit radius of 8.05 \rearth.  There is some evidence 
in the model fit for an eccentric orbit or stellar radius as large as 1.4 \rsun.
We did not detect a contaminating BGEB: alternating transit events have 
the same depth, no decrease in brightness is observed at the predicted 
occultation time, and the pixel centroid analysis yielded a clean result for 
a transit on KIC~6315331 without any detected astrometric motion. The 2MASS 
and AO images are shown in Figure \ref{fig_im_montage} (bottom, left and right).  Because this 
is the faintest of the stars (Kepler magnitude of 15.64), the AO images can only rule 
out contaminating background stars within  $\Delta M_V < 2.7$ magnitudes at separations 
larger than 0\farcs5. The AO contrast sensitivities are listed in Table \ref{tab_ao}.

\subsection{KIC 8242434}
Planet Hunters identified a single transit event in the Q1 data for KIC~8242434. When the 
Q2 data were released, two additional transit events were identified that were separated by 
44 days.  In consultation with the Kepler team, we learned that the TPS had flagged 
this star with a MES of about $10 \sigma$. Because this was a single event, the 
data verification was not processed until Q2, and was not classified as a KOI.

The KIC lists a Kepler magnitude of 13.054 and \gr\ color 
of 0.937,  \teff = 4665K, \logg = 4.176, a high metallicity of +0.437 and a stellar 
radius of 1.337 \rsun for KIC~8242434. We analyzed a Keck HIRES spectrum with 
\snr\ of about 55 and derive a similar temperature, \teff = 4757 $\pm 60$K. 
However, we find a higher surface gravity, \logg = 4.608$\pm 0.1$, consistent with a 
main sequence luminosity class. The wings of the Mg I B triplet lines 
(Figure \ref{fig_mgb_caii}) are broad and by eye are consistent with the higher surface gravity.  
Our analysis also yields a lower metallicity, \fe\ = 0.07 and \vsini = 0.4 \kse.  
The \caii\ lines (Figure \ref{fig_mgb_caii}) have emission in the line core; this emission would be
typical for a low mass main sequence star, but less common for a subgiant. 
The stellar parameters are summarized in Table \ref{tab_stellar_pars}. 

\begin{figure*}
\begin{center}
\includegraphics[angle=0, width=1.0\textwidth]{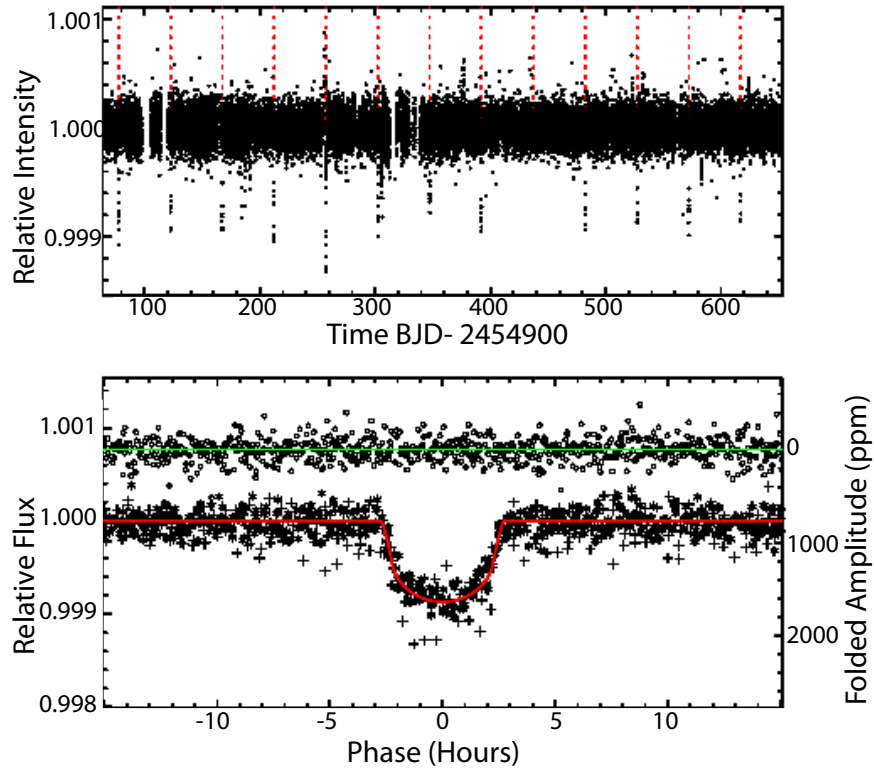}
\caption{The time series data for KIC~8242434 (top) include photometry for Q1 - Q7, 
provided by the Kepler team. Planet Hunters flagged a single transit in the Q1 data and 
two additional transits were found in the Q2 data.  The bottom panel shows the data folded 
at the prospective orbital period, 44.9634 days. Unfortunately, the pixel centroid check shows 
that this is llkely a background eclipsing binary system.}

\label{fig_8242434_lc}

\end{center}
\end{figure*}

Figure \ref{fig_8242434_lc} shows the time series and phase-folded Q1 - Q7 photometry for KIC~8242434. 
The light curve does not show evidence for a BGEB: the transit depth is constant 
for alternating transits and no dimming occurs at the predicted time of occultation in the 
phase-folded data just above the transit curve. The orbital 
period is modeled as 44.963888d. A Monte Carlo analysis was used to iterate to the
self-consistent stellar parameters listed in Table \ref{tab_stellar_pars} (again, there was not a good constraint 
for the stellar age). The stellar radius is estimated to be 
0.719 \rsun, and together with the transit depth, this implies a planet radius of 
2.32 \rearth. The parameters for the planet candidate are summarized in Table \ref{tab_planet_pars}. 

\begin{figure*}
\begin{center}
\includegraphics[angle=0, width=1.0\textwidth]{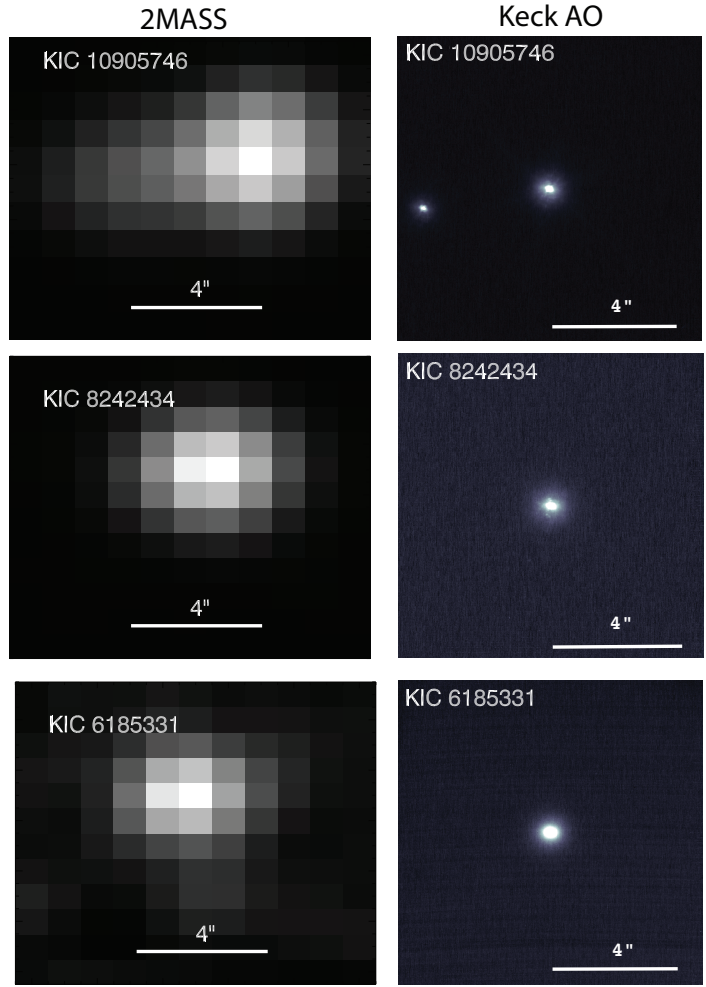}
\caption{The 2MASS K band images (left) and AO images (right) for the two planet candidate hosts, 
KIC~10905746 and KIC~6185331 and for a star where a background eclipsing binary was found, KIC~8242434. The horizontal line indicates the image scale in arcseconds. North is up in these images and East is to the left. KIC~10905746 is shown in the top panel; the 2MASS image shows distortion from a nearby star at about 4 arcseconds due East, which is completely resolved by the AO K-band image (right). 
The middle panel shows 2MASS and AO images for KIC~8242434 and to the magnitude limits listed in Table 3, no additional sources are observed, however the photocenter was observed to shift during the prospective transit, indicating
that a nearby backgroud eclipsing binary star producing the transit signal. The bottom panel shows images for KIC~6315331 with weaker limits on excluded background sources because of the intrinsic faintness of 
this star.}

\label{fig_im_montage}

\end{center}
\end{figure*}

The measured position of the transit source shows a statistically significant (5.7 sigma) 0.6 arcsec offset 
from KIC 8242434, indicating that the transit signal is likely due to a dim background binary.  The source 
position is measured by taking robust weighted average of the observed transit source position in quarters 
1-8, as determined by centroiding the difference between average in-transit and out-of-transit pixels \citep{b11}. 
Modeling indicates that this offset is not due to systemic centroid biases due, for example, to crowding.
The K-band 2MASS image is shown in Figure \ref{fig_im_montage} (middle, left) and the AO 
image (Figure \ref{fig_im_montage}, right) shows some unusually bright speckles within an arcsecond,
with the most prominent one in the south-east. The AO images and pixel centroid analysis casts doubt on 
the planet interpretation and suggests the presence of a confusing background 
source; likely a BGEB.

\subsection{KIC~11820830}
\begin{figure*}
\begin{center}
\includegraphics[angle=0, width=1.0\textwidth]{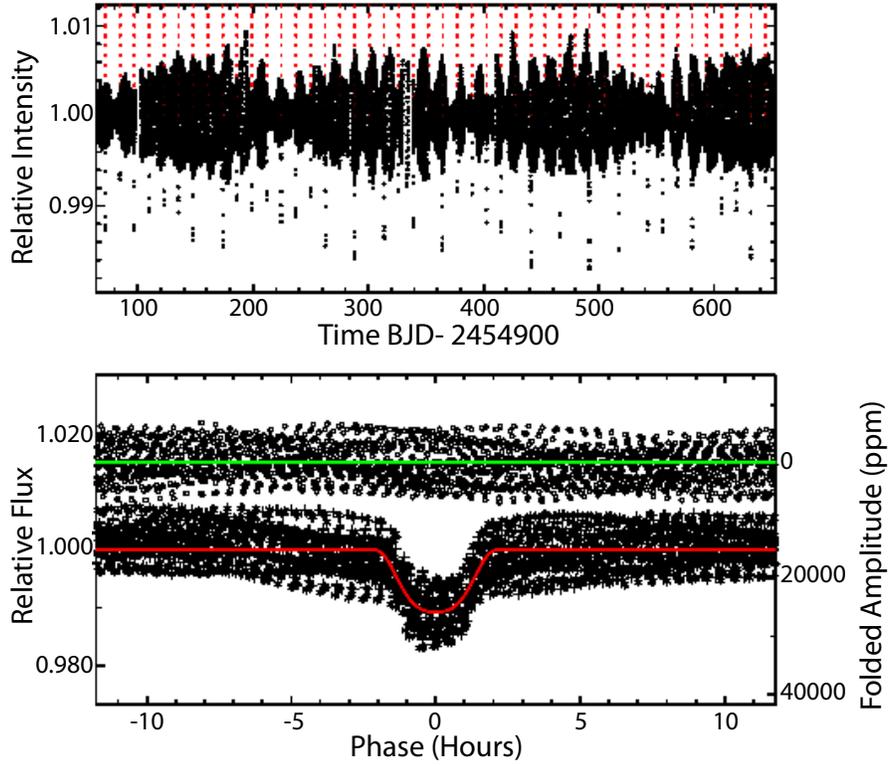}
\caption{The time series data for KIC~11820830 (top) include Q1 - Q7 data. This star has a remarkably variable background 
However, Planet Hunters were able to see past that structure and flagged several transits in the Q1 data. 
The bottom panel shows the phase-folded data with the prospective orbital period, 12.7319 days. Unfortunately,
the best model for this star suggests that the primary is an early type star with an eclipsing M or K dwarf companion}

\label{fig_11820830_lc}

\end{center}
\end{figure*}

KIC~11820830 exhibits significant oscillations, however, participants 
readily identified several transit events in the Q1 light curve.  The Kepler 
TPS had also flagged this star with a MES of $46 \sigma$, the highest 
\snr\ threshold of any of the candidates presented in this paper. However, 
the light curve failed additional tests and was not processed by the data verification 
pipeline. Figure \ref{fig_11820830_lc} shows the remarkable time series (top) and phase-folded (bottom) 
light curves for Q1 - Q7 observations of this star.  

The Kepler Input Catalog lists stellar parameters for KIC~11820830, 
including Kepler magnitude of 12.087, \gr\ color of 0.198, stellar 
radius of 1.428 \rsun, \teff = 7007K, \logg = 4.224 and [Fe/H] = -0.009. 
We obtained a spectrum of this star using HIRES on Keck with 
SNR ~ 90. We carried out spectral synthesis modeling and 
derive spectroscopic properties of the star. 

This is the brightest of the our initial Planet Hunters candidates, and normally 
it would have been possible to follow-up on this star with Doppler 
measurements to confirm the mass of the transiting object. 
However, our spectroscopic analysis revealed a high rotational 
velocity, \vsini $= 52 \pm 5$ \ks which significantly reduces the intrinsic 
radial velocity precision. Figure \ref{fig_mgb_caii} shows the Keck wavelength 
segments for the Mg I B triplet and \caii\ lines respectively, and the 
high rotational velocity is apparent from the broad stellar lines in 
these Figures. The broad spectral lines also reduce the precision of our 
derived spectral parameters.  With this caveat, we report the results 
of our analysis: \teff = 6300 $\pm 250$K, \logg = 3.6 $\pm 0.2$, 
and \fe\ = +0.26 $\pm 0.2$. 

Unfortunately, the self-consistent Monte Carlo analysis indicates that KIC~11820830 
is likely to be an eclipsing binary system, with an early type primary star eclipsed by 
a K or M dwarf in an eccentric orbit.  No astrometric 
motion was detected in the pixel centroid analysis and the AO images did not detect 
an additional source with a $\Delta M_V < 4$ magnitudes at separations of 0\farcs25. 
The AO contrast sensitivities are summarized in Table \ref{tab_ao}. 

\section{Discussion} 
The Planet Hunters website was launched to engage the public in 
front-line research by presenting light curve data from the Kepler Mission.  
This project joins a growing list of citizen science Zooniverse projects, and is the 
first to present time series data, rather than images. We debated whether the unique 
pattern recognition skills of the human brain would be able to compete with the efficient computer 
algorithms. However, we expected that citizen scientists might discover unexpected patterns 
in the data or unusual types of transits, which could then be used as feedback to further improve 
the Kepler transit search algorithms.  Citizen scientists identified some unusual objects in 
the Galaxy Zoo program, and we expected that some unpredictable and unanticipated 
discoveries and correlations might also emerge from Planet Hunters.  Automated algorithms 
and citizen science are complementary techniques and both are important to make the best 
use of the Kepler data. 

An initial assessment was made of the performance and efficiency 
of the Planet Hunters participants by counting the number of transit 
events detected among the 306 candidates announced for Q1 data 
by \citet{bor10a}.  We found that Planet Hunters flagged about two 
thirds of those transit events. The deeper transits were found more often
than the shallow transits. 

In the first month after the launch of the Planet Hunters website, more than 
forty stars were flagged as possible planet transits that were not known false positives 
(grazing binaries or blended background eclipsing binaries) or published Kepler 
candidates. Because we felt it was important to preserve the integrity of the 
Kepler planet candidates, we contacted members of the Kepler team who provided important data 
verification for our top ten candidates.  More than half of these were found 
to be false positives. 

We present the first two planet candidates, discovered by Planet Hunters  
using Q1 data: KIC~10905746 and KIC~6315331, with orbital 
periods that range from 9.88 to 49.96 days and radii ranging from 2.32 to 8.0 \rearth.  
We have carried out a Monte Carlo 
analysis for a self-consistent set of stellar parameters and analyzed the pixel 
centroid's to check for astrometric motion. We also obtained adaptive optics 
(AO) observations to eliminate background eclipsing binaries (BGEBs) with 
separations wider than $\sim0\farcs5$ and $\Delta M_V< 5$ in the infrared K-band data.  
However, the pixel centroid analysis and AO observations cannot exclude 
eclipsing binaries that are are closer than 0\farcs5 or those with wider 
separations that are more than about 5 magnitudes fainter than the 
tentative planet host stars. Because such systems could still produce the 
observed light curves, these two candidates are not confirmed planets.  

We estimate false positive probabilities (FPP) for the two candidates presented here 
following the framework presented in Morton \& Johnson (2011), which relies on 
Galactic structure and stellar population synthesis models.  We consider two possible 
false positive scenarios: chance-alignment blended eclipsing binaries and hierarchical 
triple eclipsing systems, both of which can produce signals that mimic transiting planets.  
However, given that these transits are not V-shaped, we observe no secondary eclipse, 
and pixel offset calculations and AO observations indicate that any possibly blending 
systems can only reside within a fraction of an arcsecond of the target stars, we are able 
to put strong statistical constraints on the likelihood of false positive scenarios.  Assuming 
an overall 20\% occurrence rate for planets, a planet radius function $dN/dR \sim R^{-2}$, 
and the binary and multiple system properties according to \citet{r10}, as discussed 
in more detail in Morton and Johnson (2011), we derive an FPP of only 0.3\% for 
KIC 10905746 and an FPP of 5.0\% for KIC 6185331.   
The higher FPP for KIC 6185331 is set primarily by the fact that it has a deeper transit 
and thus is more susceptible to the hierarchical blend false positive scenario, which is 
not significantly constrained by the AO observations or centroid analysis.

An obvious question is why these candidates were not identified by 
the Kepler team. One motivation for the Planet Hunters project was that there might 
be odd cases that computer algorithms might miss, but that the human brain would 
adeptly identify. In fact, we learned that all of the planet candidates presented 
here had previously been flagged by the Transit Planet Search (TPS) algorithm. However, 
two of the candidates presented here had multi-quarter light curves that did not converge 
and the third candidate was dropped after Q1 because it was thought to be an evolved star. 
Therefore, these stars were not promoted to the status of a Kepler Object of 
Interest (KOI), which would have triggered extensive follow-up.  It is not really 
surprising that a few candidates failed to converge in the analysis pipelines and 
remained behind to be gleaned by Planet Hunters.  The discoveries presented in 
this paper show the challenges of field confusion for transiting planets, yet also shows 
that Citizen Scientists can make important contributions.

Planet Hunters is a novel and complementary technique to the Kepler TeamÕs 
detection algorithms with different systematics and intrinsic biases than computer based algorithms. 
Algorithms are now being developed to process Planet Hunters classifications and assess the capabilities of individual volunteers based on light curves injected with synthetic short-period planet transits. Weightings will be assigned to individuals, and an iterative process will be used to converge on final classifications for each star. These algorithms will extract transit candidates automatically, and this analysis will be presented in a future paper.

\section*{Acknowledgements}

The data presented in this paper are the result of the efforts of the Planet Hunters volunteers, without whom this work 
would not have been possible. The following list of people flagged transit events for the light curves discussed in this 
paper: Juan Camilo Arango Alvarez, Mary-Helen Armour, Ferdinand de Antoni, Frank Barnet, bhugh, Carolyn Bol, Les Bruffell,Dr. David M. Bundy,Troy Campbell, Elisabeth Chaghafi,Amirouche Chahour,Arunangshu Chakrabarty, Mathias Chapuis,Fabrice Cordary, Daniel, ClŽmentDoyen, Graham Dungworth, Michael Richard Eaton, David Evans, Raymond Ashley Evans,Evgeniy,Enrique Ferreyra, Marc Fiedler,Dave Fischer, Fin J. R. FitzPatrick,Dr Ed Foley,Lorenzo Fortunato,Sebastian Frehmel, Robert Gary Gagliano, Fabio Cesar Gozzo, Howard Hallmark,Dave Henderson,Inizan, irishcoffee, Thomas Lee Jacobs, Marta Ka\l u\.{z}na,Bill Kandiliotis,Rafal Kurianowicz,Lukasz Kurzysz,Piotr Laczny, David M. Lindberg,Janet Lomas, Luis Miguel Moreira Calado Lopes, Mihael Lujanac, Lukasz, C‡ssia Solange Lyra,Jacek M,Riccardo Marzi, Karen McAuley,Tomasz Miller, Cedric MOULIS, Adrian Nicolae,Njaal,Michael W Novak, Osciboj, Kai Pietarinen, Anna Podjaska, Marc A. Powell (Omeganon), Gerry A. Prentice,John M. Rasor,RenŽ-Pierre BUIGUES,Andres Eloy Martinez Rojas,RouzŽ,Andrey Sapronov, Matt Schickele, Terrance B. Schmidt, David Smith,Paulina Sowick, Lubom\'ir \`Stiak, Charles H Tidwell III, tuckdydes, v5anw, Joop Vanderheiden,VIATGŽ, vovcik91, SŽbastien Wertz, Bohdan Widla, Steven C Wooding, Charles Yule.

DF acknowledges funding support from Yale University and support from the 
NASA Supplemental Outreach Award, 10-OUTRCH.210-0001. DF thanks the Yale Keck TAC for telescope 
time used to obtain data for this paper.  MES is supported by an NSF Astronomy and Astrophysics Postdoctoral Fellowship under award AST-100325.  Support for the work of KS was provided by NASA through 
Einstein Postdoctoral Fellowship grant numbers PF9-00069, issued by the Chandra X-ray Observatory Center, 
which is operated by the Smithsonian Astrophysical Observatory for and on behalf of NASA under contract 
NAS8-03060.The Zooniverse is supported by The Leverhulme Trust. 
We gratefully acknowledge the dedication and achievements of Kepler Science Team 
and all those who contributed to the success of the mission. We acknowledge use of public release data 
served by the NASA/IPAC/NExScI Star and Exoplanet Database, which is operated by the Jet Propulsion 
Laboratory, California Institute of Technology, under contract with the National Aeronautics and Space Administration.  
We particularly thank the organizers (Charles Beichman, Dawn Gelino, Carolyn Brinkman) and lecturers 
(David Ciardi, Stephen Kane, Kaspar von Braun) at the July 2010 Sagan Summer Workshop for providing 
information and guidance that led to the inspiration for the Planet Hunters site.  The Kepler public release 
data is primarily hosted by the Multi-mission Archive (MAST) at the Space Telescope Science Institute 
(STScI) operated by the Association of Universities for Research in Astronomy, Inc., under NASA 
contract NAS5-26555. Support for MAST for non-HST data is provided by the NASA Office 
of Space Science via grant NNX09AF08G.  This research has made use of NASA's Astrophysics Data System 
Bibliographic Services.

\bibliographystyle{mn}
%\bibliography{bibliography}

\bsp

%Figure 1     

%Figure 2  

% Fig 3

% Fig 4

%Fig 5

%Fig 6

%Fig7

\clearpage

% table 1 (NB Updated from original)
\begin{deluxetable}{ll}
\tablenum{1}
\tablecaption{False Positive Planet Candidates\label{tab_failed}}
\tablewidth{0pt}
\tablehead{
\colhead{Starname}  & \colhead{Comments}  \\}
\startdata
KIC~11904734       &   V-shaped transit and very large radius (EB)   \\
KIC~8043052         &    Secondary occultations (EB)  \\
KIC~12009347       &    Secondary occultations (EB) \\
KIC~4913000        &    Astrometric motion in pixel centroids (BGEB)  \\
KIC~9097892         &  Astrometric motion in pixel centroids (BGEB)    \\
KIC~11820830        &   Eclipsing binary (based on model fits) \\
KIC~8242434            & Astrometric motion in pixel centroids (BGEB)  \\
\enddata
\end{deluxetable}
\clearpage

%table 2
\begin{deluxetable}{lllll}
\tablenum{2}
\tablecaption{Stellar Parameters\label{tab_stellar_pars}}
\tablewidth{0pt}
\tablehead{
\colhead{Parameter} & \colhead{10905746} & \colhead{6185331} & \colhead{8242434} & \colhead{11820830} \\
}
\startdata
Right Ascension         		& 18 54 30.92 	         & 18 57 05.75  	      & 19 39 49.22         & 19 40 51.98   \\
Declination             			& 48 23 27.6    		& 41 32 06.1    	      & 44 08 59.3           & 50 05 03.58   \\
Kepler mag          			& 13.49           		& 15.64       	      & 13.05         	  & 12.09   	   \\
\gr\                   				& 0.949           	         & 0.556     	               & 0.937                    & 0.198   	   \\
$M_{*}$ [\msun]      			& 0.578 (0.032)         & 1.027 (0.042)      & 0.761 (0.028)   	   & 2.25 (0.3)    	  \\
$R_{*}$ [\rsun]				& 0.548 (0.026)         & 1.27 (0.17)  	      & 0.719 (0.031) 	   & 4.1 (0.3)     	  \\
Z						& 0.0119 (0.003)       & 0.0261 (0.0032) & 0.0234 (0.003)	   &			  \\	
Age [Gyr]					& \nodata			& 8.7 (1.5)		      & \nodata		   &			  \\
$L_{*}$ [\lsun]      			& 0.086 (0.081)  	& 1.02 (0.03)          & 0.77 (0.04)  	   & 2.25 (0.3)    	  \\
$\rho_{*}$ [$\rm{g cm^-3}$]	& 4.97 (0.54)     	& 0.70 (0.26)  	     & 2.9 (0.38)	            &			  \\
\teff  [K]          				& 4240 (112)   	         & 5619 (80)   	     & 4757 (60)     	  & 6300 (250)    \\
\fe\         					& -0.23 (0.1)    		& +0.11 (0.15)  	     & +0.07 (0.08)        & +0.26 (0.2) 	 \\
\vsini [\kse]          			& 1.1 (0.50)    		& 0.5 (0.50)            & 0.4 (0.50)             & 52 (5)  	    	 \\
\logg                  				& 4.724 (0.028)         & 4.239 (0.098)    & 4.608 (0.041)  	 & 3.6 (0.2)   	 \\
\enddata
\end{deluxetable}
\clearpage

% table 3
\begin{deluxetable}{llllll}
\tablenum{3}
\tablecaption{$\Delta$ K Magnitude AO Exclusion Limits\label{tab_ao}}
\tablewidth{0pt}
\tablehead{
\colhead{Starname}  & \colhead{0\farcs25} & \colhead{0\farcs5}  & \colhead{1\farcs0}  & \colhead{2\farcs0} & \colhead{4\farcs0} \\}
\startdata
KIC~10905746        &   3.7                         &  5.6                                &  7.8                         &  8.6                             &  8.6          \\
KIC~6185331          &   1.1                          &  2.7                                &  4.8                         &  5.2                             &  5.3          \\
KIC~8242434          &   3.5                          &  5.1                                &  7.2                         &  7.8                             &  7.9          \\
KIC~11820830        &   4.2                          &  5.7                                &  7.2                         &  7.6                             &  7.6          \\
\enddata
\end{deluxetable}
\clearpage

%table 4
\begin{deluxetable}{llll}
\tablenum{4}
\tablecaption{Characteristics of Planet Candidates\label{tab_planet_pars}}
\tablewidth{0pt}
\tablehead{\colhead{Parameter}  & \colhead{10905746}  & \colhead{6185331}   \\
}
\startdata
T$_0$ [BJD-2454900]      	& 71.4045 (0.0102)    	& 92.9877 (0.0028)   	 	\\
Orb. Per. [d]       		& 9.8844 (0.0087) 	  	& 49.76971 (0.00039) 		\\
Impact parameter, b		& 0.82 (0.21)			& 0.642 (0.142)			         \\
$R_{PL} / R_*$			& 0.0442 (0.0110)		& 0.0581 (0.0018)	 		\\
$e sin\omega$			& 0.08 (0.42)			& 0.10 (0.32)			         \\
$e cos\omega$			& 0.00 (0.43)			& 0.00 (0.34)				\\
$R_{PL}$ [\rearth]    		& 2.65 (0.67)			& 8.05 (1.08)          		         \\
Incl [deg]         			& 88.42(0.42)           	    	& 89.20 (0.21)    			\\
$a / R_*$				& 29.4 (1.1)			& 38.1 (8.4)				\\
a [AU]				& 0.0751 (0.0014)   		& 0.2672 (0.0036)			\\
T depth (ppm)			& 1881. (343.)			& 3633. (59.)			         \\
\enddata
\end{deluxetable}
\clearpage

\label{lastpage}

\end{document}